\documentclass[a4paper,draft,reqno,12pt]{amsart}
\usepackage{amsmath,amssymb,eucal,bbold}
\usepackage{latexsym}
\usepackage{array}
\DeclareMathOperator{\Tr}{\mathrm{Tr}}
\DeclareMathOperator{\Det}{\mathrm{Det}}
\renewcommand{\d}{\partial}
\renewcommand{\Sp}{\mathrm{Sp}}
\renewcommand{\sp}{\mathfrak{sp}}
\newcommand{\1}{\mathbb{1}}
\newcommand{\CC}{\mathbb{C}}
\newcommand{\EE}{\mathbb{E}}
\newcommand{\HH}{\mathbb{H}}
\newcommand{\MM}{\mathbb{M}}
\newcommand{\RR}{\mathbb{R}}
\newcommand{\eA}{\mathcal{A}}
\newcommand{\eC}{\mathcal{C}}
\newcommand{\eG}{\mathcal{G}}
\newcommand{\eI}{\mathcal{I}}
\newcommand{\eIt}{\Hat{\mathcal{I}}}
\newcommand{\eJ}{\mathcal{J}}
\newcommand{\eL}{\mathcal{L}}
\newcommand{\eM}{\mathcal{M}}
\newcommand{\eO}{\mathcal{O}}
\newcommand{\eS}{\mathcal{S}}
\newcommand{\fg}{\mathfrak{g}}
\newcommand{\so}{\mathfrak{so}}
\newcommand{\Cl}{\mathrm{C}\ell}
\newcommand{\Mat}{\mathrm{Mat}}
\newcommand{\SL}{\mathrm{SL}}
\newcommand{\SO}{\mathrm{SO}}
\newcommand{\Spin}{\mathrm{Spin}}

\newcommand{\sA}{\mathsf{A}}
\newcommand{\sB}{\mathsf{B}}
\newcommand{\sC}{\mathsf{C}}
\newcommand{\sK}{\mathsf{K}}
\newcommand{\sP}{\mathsf{P}}
\newcommand{\yms}{\mathsf{SYM}_{9{+}1}}

\newcommand{\eLYM}{\mathcal{L}_{\mathsf{YM}}}
\newtheorem{lem}{Lemma}

\begin{document}

\title[super Yang--Mills, instantons and triholomorphic
curves]{Supersymmetric Yang--Mills, octonionic instantons and
triholomorphic curves}
\author[Figueroa-O'Farrill]{JM Figueroa-O'Farrill}
\author[K\"ohl]{C K\"ohl}
\author[Spence]{B Spence}
\thanks{Supported by the EPSRC under contract GR/K57824.}
\thanks{Supported by an EPSRC postgraduate studentship.}
\thanks{EPSRC Advanced Fellow.}
\address{\begin{flushright}Department of Physics\\
Queen Mary and Westfield College\\
University of London\\
London E1 4NS, UK\end{flushright}}
\email{j.m.figueroa@qmw.ac.uk}
\email{c.koehl@qmw.ac.uk}
\email{b.spence@qmw.ac.uk}
\begin{abstract}
In four-dimensional gauge theory there exists a well-known
correspondence between instantons and holomorphic curves, and a
similar correspondence exists between certain octonionic instantons
and triholomorphic curves.  We prove that this latter correspondence
stems from the dynamics of various dimensional reductions of
ten-dimensional supersymmetric Yang--Mills theory.  More precisely we
show that the dimensional reduction of the ($5{+}1$)-dimensional
supersymmetric sigma model with hyperk\"ahler (but otherwise
arbitrary) target $X$ to a four-dimensional hyperk\"ahler manifold $M$
is a topological sigma model localising on the space of triholomorphic
maps $M\to X$ (or hyperinstantons).  When $X$ is the moduli space
$\eM_K$ of instantons on a four-dimensional hyperk\"ahler manifold
$K$, this theory has an interpretation in terms of supersymmetric
gauge theory.  In this case, the topological sigma model can be
understood as an adiabatic limit of the dimensional reduction of
ten-dimensional supersymmetric Yang--Mills on the eight-dimensional
manifold $M\times K$ of holonomy $\Sp(1)\times \Sp(1) \subset
\Spin(7)$, which is a cohomological theory localising on the moduli
space of octonionic instantons.
\end{abstract}
\maketitle

\section{Introduction}

There is increasing evidence of an underlying unifying structure in
superstring theory, M-theory. This is based on the continuing progress
of research on duality.  Perhaps the boldest proposal yet to have
emerged from this duality revolution is the Matrix conjecture
\cite{Matrix}, which states that M-theory can be described (at least
in the infinite momentum frame) by the dimensional reduction to one
dimension of 10-di\-men\-sion\-al supersymmetric Yang--Mills ($\yms$)
in the limit in which the rank of the gauge group goes to infinity.
Despite some recent successes, it is fair to say that we are still far
from understanding M-theory compactifications from a Matrix-theoretic
point of view; but the results on toroidal compactifications seem to
suggest that the dimensional reductions of $\yms$ play an important
r\^ole.

In this paper, we report the results of our continuing exploration of
dimensional reductions of $\yms$ on curved riemannian manifolds.
Reductions on manifolds of special holonomy give cohomological field
theories which localise on finite-dimensional moduli spaces.  By
taking certain geometric limits one obtains interesting
relations involving different moduli spaces.  In the particular
example we treat here, we relate the moduli space of octonionic
instantons in eight dimensions in an adiabatic limit to triholomorphic
curves, generalising the well-known relationship in four dimensions
between instantons and holomorphic curves.  This rather rich interplay
between solitonic physics and complex geometry is thus seen to be a
consequence of the underlying relationship with $\yms$.

Whilst we do not explore here in detail the M-theory applications and
understanding of these results, it is clear that these will be based
upon an analysis of wrapped and intersecting D-branes.  As
perturbative string propagation constrains the local geometry of the
spacetime, so D-branes probe the geometry of submanifolds, and in
particular those of minimal volume.  A very important class of
examples of minimal submanifolds are the calibrated geometries
introduced in \cite{HL} and rediscovered in string theory as
supersymmetric cycles \cite{BBS}.  It is thus not a coincidence that
such geometries are richest in those Ricci-flat manifolds possessing
parallel spinors; that is, supersymmetric vacua.  These include the
manifolds of reduced holonomy which we study here. The relation
between spinors and calibrations is of course well established
\cite{H,LM}; but many of the applications to topology seem to be
novel.

When a D-brane wraps around a supersymmetric cycle, it gives rise to a
solitonic state of the effective compactified theory.  In fact, much
of the solitonic spectrum can be realised in this fashion.  The mass
of the soliton is given in the simplest cases by the volume of the
cycle and the fact that the cycle is minimal is the geometric
restatement of the BPS condition.  On the other hand we may focus on
the effective theory on the worldvolume of the D-brane, namely the
dimensional reduction of $\yms$.  The interesting phenomenon is that
for curved euclidean branes these theories are cohomological: the BRST
symmetry being essentially a supersymmetry with a parallel spinor as
parameter.  In this fashion one can recover most known cohomological
theories both in dimension $d\leq4$ \cite{BSV,BTNT2too} and in higher
dimension \cite{BTESYM,AFOS,FIM}. In particular this makes contact
with the recent generalisations of Donaldson--Floer--Witten theory to
higher dimensions \cite{DT,BKS,BKS2,AOS}.

This programme has thus far focused on supersymmetric gauge theories
and it is natural to ask if other cohomological field theories may be
understood in this manner. In this paper we will show that the
four-dimensional topological sigma model of Anselmi and Fr\`e
\cite{AF} may indeed also be understood, for some choices of target
space, as arising from wrapped euclidean D-branes.  This observation
has a remarkable consequence: it allows us to understand a gauge
theoretic result mentioned in \cite{DT,Thomas} relating the moduli
space of certain octonionic instantons in eight dimensions and
triholomorphic curves, similar to the well-known relation between
four-dimensional instantons and holomorphic curves
\cite{DS1,DS2,BJSV}.

This paper is organised as follows.  We start in Section 2 by
considering the six-dimensional supersymmetric sigma model with a
hyperk\"ahler target $X$.  In Section 3 we dimensionally reduce this
model in one space and one time dimensions to four {\em euclidean\/}
dimensions.  In Section 4 we consider the theory on a curved
four-dimensional manifold $M$. Only when $M$ is also hyperk\"ahler do
we have any supersymmetry left over.  The remaining supercharges are
BRST-like and one them can be interpreted as the BRST operator in a
topological sigma model localising on the moduli space of
triholomorphic maps $M \to X$, or hyperinstantons.  This is the
hyperk\"ahler version of Gromov--Witten theory, discussed by Anselmi
and Fr\`e \cite{AF} in a different context.  In Section 5 we
specialise the sigma model to the case in which $X = \eM_K$ is the
moduli space of Yang--Mills instantons on a hyperk\"ahler
four-dimensional manifold $K$ (e.g., a $K3$ surface), which is
well-known to be a hyperk\"ahler manifold itself.  In this case we
show that the topological sigma model can also be obtained by
dimensionally reducing $\yms$ on the eight-dimensional manifold $M
\times K$ in the limit in which $K$ becomes small.  This reformulation
sets up an isomorphism between triholomorphic curves in the moduli
space $\eM_K$ and a certain moduli space of octonionic Yang--Mills
instantons.  We close the paper with four technical appendices.

\subsubsection*{A foreword on notation}

Our conventions will be as follows: throughout the paper we will use
the notation $\MM^{s{+}t}$ to refer to ($s{+}t$)-dimensional Minkowski
space.  In addition $\EE^d = \MM^{d{+}0}$ will denote $d$-dimensional
euclidean space.  Spinor notation will follow
\cite{LM}.  In particular, $\Cl(s,t)$ denotes the Clifford algebra
\begin{equation*}
\Gamma_\mu \Gamma_\nu + \Gamma_\nu \Gamma_\mu = -2 \eta_{\mu\nu} \1~,
\end{equation*}
where $\eta_{\mu\nu}$ is diagonal with signature $+s{-}t$.  Our
notation for representations of the spin groups is the following.  The
trivial, vector and adjoint representations are denoted $\bigwedge^0$,
$\bigwedge^1$, and $\bigwedge^2$ respectively.  The half-spin
representations are denoted $\Delta$ for odd-dimensional spin groups
and $\Delta_\pm$ for the even-dimensional spin groups.  We will make
use of the notation $[\rho]$ to denote the underlying real
representation of a complex representation $\rho$ with a real
structure.  In other words, $\rho \cong [\rho]\otimes\CC$.  Other
group theory notation will be introduced as needed.  Our conventions
for $\epsilon_{AB}$ are as follows.  We take $\epsilon^{12} =
-\epsilon^{21} = \epsilon_{12} = - \epsilon_{21} = +1$.  We raise and
lower indices using the ``northeast'' convention:
\begin{equation*}
O_A = \epsilon_{AB} O^B \qquad\text{and}\qquad
O^A = O_B \epsilon^{BA} = - \epsilon^{AB} O_B~.
\end{equation*}
Therefore ${\epsilon_A}^B = \epsilon_{AC} \epsilon^{CB} = -
\delta_A^B$, and hence $\epsilon_{AB} \epsilon^{AB} = 2$.

\section{The six-dimensional supersymmetric sigma model}

In this section we will review the supersymmetric sigma model in
$\MM^{5{+}1}$.  The bosons in the supersymmetry sigma model are maps
from $\MM^{5{+}1}$ to a hyperk\"ahler manifold $X$ of (real) dimension
$4n$.  Therefore there are $4n$ real bosonic degrees of freedom, both
on-shell as off-shell.  Normally the fermions in the sigma model would
take values in (the pull-back of) the tangent bundle $TX$ of the
target manifold.  However in this case, this prescription does not
give rise to a match between the bosonic and fermionic degrees of
freedom.  Because $\Spin(5,1)$ is isomorphic to $\SL(2,\HH)$, the
smallest spinor in $\MM^{5{+}1}$ is a Weyl spinor which has 8 real
components.  $4n$ such spinors would have $32n$ real components,
reduced to $16n$ on-shell.  We need to cut the dimension by a factor
of 4.  This is accomplished in a manner we now detail.

\subsection{Some hyperk\"ahler geometry}

The complexified tangent bundle $T_\CC X$ of a hyperk\"ahler manifold
decomposes under the maximal subgroup $\Sp(n) \cdot \Sp(1) \subset
\SO(4n)$ as $T_\CC X \cong V\otimes\Sigma$, where $\Sigma$ is a
complex two-dimensional $\Sp(1)$ bundle and $V$ is a complex
$2n$-dimensional $\Sp(n)$ bundle.  These are associated to the
fundamental representations $\lambda^1$ of $\Sp(n)$ and $\sigma^1$ of
$\Sp(1)$.  Indeed, under $\Sp(n)\cdot\Sp(1) \subset \SO(4n)$, the
vector representation of $\SO(4n)$ obeys $\bigwedge^1 \cong [\lambda^1
\otimes \sigma^1]$.

The holonomy being $\Sp(n)$ means that the above decomposition of
$T_\CC X$ is preserved under parallel transport and, in addition, that
$\Sigma$ is a trivial bundle. The canonical real structure of $T_\CC
X$ is the product of the natural quaternionic structures in $\Sigma$
and $V$.  Because the spinor representations $\Delta_\pm$ of
$\Spin(5,1)$ also possess a quaternionic structure, the tensor
products $\Delta_\pm\otimes \lambda^1$ and $\Delta_\pm\otimes\sigma^1$
possess real structures.  Therefore we will be able to impose reality
conditions on the fermions and on the supersymmetry parameters
respectively.\footnote{Such spinors are known as symplectic
Majorana--Weyl spinors, and they exist in spacetimes of signature
$(s,t)$ with $s-t=4\mod 8$.}  The representation $\Delta_\pm\otimes
\lambda^1$ is complex $8n$-dimensional.  The reality condition leaves
$8n$ real components which gives $4n$ physical degrees of freedom for
the spinors, matching the number of bosonic physical degrees of
freedom.  Similarly $\Delta_\pm\otimes\sigma^1$ is complex
eight-dimensional and the reality condition leaves the expected 8 real
components of the supercharge.

We now introduce some notation to describe the fields in the sigma
model.  First we have $4n$ bosons $\phi^i$ which are coordinates of
the target manifold.  The isomorphism $T_\CC X \cong V\otimes\Sigma$
is given explicitly by objects $\gamma^i_{A a}$.  Here $A,B,\ldots$
are $\Sp(1)$ indices associated with the representation $\sigma^1$ and
take the values 1 and 2, whereas $a,b,\ldots$ are $\Sp(n)$ indices
associated with $\lambda^1$ and run from 1 to $2n$.  The bundle
$\Sigma$ being trivial allows a constant $\Sp(1)$-invariant symplectic
form $\epsilon_{AB}$; whereas $V$ admits an $\Sp(n)$-invariant
symplectic form $\omega_{ab}$.  In terms of these symplectic forms,
the metric $g_{ij}$ on $X$ can be written as:
\begin{equation}\label{eq:metric}
g_{ij} \gamma^i_{A a} \gamma^j_{B b} = \epsilon_{AB} \omega_{ab}~.
\end{equation}
Because the holonomy lies in $\Sp(n)$, not just the metric $g$, but
also the symplectic forms $\epsilon$ and $\omega$ are parallel; whence
so are the maps $\gamma^i_{A a}$.  We choose to trivialise $\Sigma$
globally and put on it the zero connection.  This way any constant
section is parallel.

We now record some identities that will prove useful in the sequel.
The maps $\gamma^i_{A a}$ being parallel imply
\begin{equation}\label{eq:gcc}
\d_j\gamma^i_{Aa} = \Hat\Gamma_j{}^b{}_a \gamma^i_{Ab} -
\Gamma_j{}^i{}_k \gamma^k_{Aa}~,
\end{equation}
where $\Hat\Gamma_j{}^b{}_a$ is the reduction to $\Sp(n)$ of the
riemannian connection on $X$.  Using this equation and equation
\eqref{eq:metric} one deduces that
\begin{equation*}
\Hat\Gamma_i{}^b{}_a = \tfrac12 \gamma^{Ab}_k \d_i \gamma^k_{Aa} +
\tfrac12 \gamma^{Ab}_k \Gamma_i{}^k{}_j \gamma_{Aa}^j~.
\end{equation*}
We define the $\Sp(n)$ curvature by
\begin{equation*}
\Hat R_{ij}{}^b{}_c \equiv \d_i\Hat\Gamma_j{}^b{}_c -
\d_j\Hat\Gamma_i{}^b{}_c + \Hat\Gamma_i{}^b{}_d\Hat\Gamma_j{}^d{}_c -
\Hat\Gamma_j{}^b{}_d\Hat\Gamma_i{}^d{}_c~;
\end{equation*}
whence
\begin{equation}\label{eq:spncurv}
\Hat R_{ij}{}^b{}_a = \tfrac12 R_{ij}{}^k{}_\ell\,
\gamma^{Ab}_k\gamma_{Aa}^\ell~.
\end{equation}
The hyperk\"ahler curvature tensor is the totally symmetric tensor
$\Omega_{abcd}$ defined by:
\begin{equation}\label{eq:hkcurv}
\Omega_{abcd} = \tfrac14 \epsilon^{AB}\epsilon^{CD} \gamma^i_{A a}
\gamma^j_{B b} \gamma^k_{C c} \gamma^\ell_{D d} R_{ijk\ell}~.
\end{equation}
Therefore using \eqref{eq:spncurv} one finds that
\begin{equation}\label{eq:curvs}
{\Hat R}_{ijab} \gamma^i_{Cc} \gamma^j_{Dd} =
\Omega_{abcd}\epsilon_{CD}~.
\end{equation}

\subsection{The action and the supersymmetry algebra}

As a real algebra, the Clifford algebra $\Cl(5,1)$ is isomorphic to
$\Mat_4(\HH)$.  This means that the irreducible representation of
$\Cl(5,1)$ is a complex eight-dimensional representation with a
quaternionic structure.  The $\Hat\Gamma_M$, for $M=0,1,\ldots,5$, are
then complex $8\times 8$ matrices.  The charge conjugation matrix
$\Hat\sC$ obeys $\Hat\sC^t = - \Hat\sC$ and $\Hat\Gamma_M^t = +
\Hat\sC \Hat\Gamma_M \Hat\sC^{-1}$.  The chirality operator
$\Hat\Gamma_7 \equiv \Hat\Gamma_0 \Hat\Gamma_1 \cdots
\Hat\Gamma_5$ obeys $\Hat\Gamma_7^2 = + \1$.  This means that the
representations $\Delta_\pm$ of $\Spin(5,1)$ will inherit the
quaternionic structure.

We can now write down the following lagrangian (see Appendix
\ref{sec:ssm} for the derivation)
\begin{multline}\label{eq:ssm}
\eL = \tfrac{1}{2} g_{ij} \d_M \phi^i \d^M \phi^j + \tfrac{1}{2}
\omega_{ab} \Bar\Psi^a \Hat\Gamma^M D_M \Psi^b\\
- \tfrac1{48} \Omega_{abcd} (\Bar\Psi^a \Hat\Gamma_M \Psi^b)
  (\Bar\Psi^c \Hat\Gamma^M\Psi^d)\,.
\end{multline}

In this expression, the $\Psi^a$ are positive-chirality spinors
satisfying a symplectic Majorana condition spelled out in \cite{FKS}.
For the present purposes it is enough to think of $\Psi^a$ as
transforming under the representation $[\Delta_+\otimes\lambda^1]$ of
$\Spin(5,1)\times \Sp(n)$.  It should be emphasised, however, that
these symmetries are of a different kind: $\Spin(5,1)$ is a global
symmetry of the theory, whereas $\Sp(n)$ is local: being what
diffeomorphisms of $X$ induce once the tangent bundle has been reduced
to $\Sp(n)$.  The conjugate $\Bar\Psi^a$ is given by $\Bar\Psi^a =
(\Psi^a)^t \Hat\sC$.  Finally, the covariant derivative is given by
\begin{equation*}
D_M \Psi^a = \d_M \Psi^a + \Hat\Gamma_i{}^a{}_b \d_M\phi^i \Psi^b~.
\end{equation*}

The above lagrangian is invariant under supersymmetry transformations:
\begin{align}\label{eq:6dsusy}
\delta_\varepsilon \phi^i &= \gamma^i_{A a}
\Bar\varepsilon^A\Psi^a\notag\\
\delta_\varepsilon \Psi^a &= \gamma_i^{A a} \Hat\Gamma^M \d_M \phi^i
\varepsilon_A - \Hat\Gamma_i{}^a{}_b \delta_\varepsilon\phi^i
\Psi^b~,\notag\\
\implies  \delta_\varepsilon \Bar\Psi^a &= \gamma_i^{A a} \d_M
\phi^i \Bar\varepsilon_A \Hat\Gamma^M - \Hat\Gamma_i{}^a{}_b
\delta_\varepsilon\phi^i \Bar\Psi^b~,
\end{align}
where $\varepsilon^A$ is a constant {\em negative\/}-chirality Weyl
spinor with values in $\Sigma$ also subject to a symplectic Majorana
condition (see \cite{FKS}) which we summarise simply by saying that
$\varepsilon^A$ transforms according to the representation
$[\Delta_-\otimes\sigma^1]$ of $\Spin(5,1)\times\Sp(1)$.  Therefore,
as for $\Psi^a$, $\Bar\varepsilon^A = \left(\varepsilon^A\right)^t
\Hat\sC$.  The Noether current generating the supersymmetry is given
by
\begin{equation*}
S^{AM} = \omega_{ab} \gamma_i^{Ab}
\d_N\phi^i\Hat\Gamma^N\Hat\Gamma^M\Psi^a~.
\end{equation*}

\section{Dimensional reduction to four dimensions}

In this section we describe the dimensional reduction of the
supersymmetric sigma model from $\MM^{5{+}1}$ to $\EE^4$ and its
extension to a four-dimensional hyperk\"ahler manifold.

\subsection{Properties of spinors}\label{sec:4dspinors}

The supersymmetric sigma model described by equation \eqref{eq:ssm} is
invariant under the Lorentz group $\SO(5,1)$.  Upon reduction to
$\EE^4$, this symmetry breaks down to $\SO(4) \times \SO(1,1)$.  The
Clifford algebra $\Cl(5,1)$ is isomorphic to $\Cl(4,0) \otimes
\Cl(1,1)$ and we can choose our $\Gamma$-matrices in a way that they
reflect this isomorphism:
\begin{equation*}
\Hat\Gamma_m = \Gamma_m\otimes \sigma_3\qquad
\Hat\Gamma_0 = \1\otimes \sigma_1\qquad\text{and}\qquad
\Hat\Gamma_5 = \1\otimes (-i\sigma_2)~,
\end{equation*}
where $m$ runs from 1 to 4, and $\Gamma_m$ are the generators of
$\Cl(4,0)$.  The chirality operator $\Hat\Gamma_7$ in such a
representation is given by $\Gamma_5 \otimes \sigma_3$.  Therefore the
positive chirality spinor $\Psi^a$ can be written as $\Psi^a =
(\psi^a_+~\psi^a_-)^t$, where $\Gamma_5 \psi^a_\pm = \pm
\psi^a_\pm$. Similarly the supersymmetry parameter $\varepsilon^A =
(\varepsilon^A_-~\varepsilon^A_+)^t$, with $\pm$ again referring to
the four-dimensional chirality.  The charge conjugation matrix $\Hat\sC$
is given by $\sC\otimes\sigma_1$, where $\sC$ is the charge
conjugation matrix in $\EE^4$ and satisfies $\sC^t = - \sC$ and
$\Gamma_m^t = - \sC \Gamma_m \sC^{-1}$.  We can organise our fields in
terms of representations of $\Sp(n) \times \Spin(4) \times \SO(1,1)$.
In fact, it is more convenient to consider the following group $G =
\Sp(n) \times \Sp(1)_T \times \Sp(1)_+ \times \Sp(1)_- \times
\SO(1,1)$, where the maximal subgroup $\Sp(n)\times \Sp(1)_T \subset
\Spin(4n)$ is the symmetry of the target space $X$, and
$\Sp(1)_+\times \Sp(1)_- \times \SO(1,1)$ is what is left of the
Lorentz group $\SO(5,1)$ after dimensional reduction.  This is
convenient bookkeeping, but it has to be kept in mind that the above
groups play different r\^oles.  For example, $\Sp(n)$ is a gauge
symmetry -- essentially diffeomorphisms of the target space once the
group of the frame bundle has been reduced to $\Sp(n)$; $\Sp(1)_T$,
the centraliser of $\Sp(n)$ in $\SO(4n)$, is a rigid symmetry of the
target space; whereas $\Sp(1)_\pm$ and $\SO(1,1)$ are rigid symmetries
of the `spacetime'.  The fields and parameters transform as follows
under $G$:
\begin{align*}
\phi^i &\qquad [\lambda^1 \otimes \sigma^1 \otimes 1 \otimes 1]_0\\
\psi^a_+ &\qquad [\lambda^1 \otimes 1 \otimes \sigma^1 \otimes
1]_{+1}\\
\psi^a_- &\qquad [\lambda^1 \otimes 1 \otimes 1 \otimes
\sigma^1]_{-1}\\
\varepsilon^A_+ &\qquad [1 \otimes \sigma^1 \otimes \sigma^1 \otimes
1]_{-1}\\
\varepsilon^A_- &\qquad [1 \otimes \sigma^1 \otimes 1 \otimes
\sigma^1]_{+1}~,
\end{align*}
where the notation conforms with the one introduced earlier, $1$
denotes the trivial one-dimensional representation, and the subscript
denotes the charge of the corresponding real one-dimensional
representation of $\SO(1,1)$.  In all the cases considered above, the
real structures alluded to by the square brackets are obtained from
the real and quaternionic structures of the factors.

\subsection{Dimensional Reduction}

It is now time to write down the action in terms of the new fields.
We are setting $\d_0 = \d_5 = 0$.  This means that also $D_0 = D_5 =
0$.  The bosonic term of the action \eqref{eq:ssm} becomes
\begin{equation*}
\eL_B = \tfrac12 g_{ij} \d_m \phi^i \d_m\phi^j~.
\end{equation*}
In order to write down the fermionic terms, we notice that $\Bar\Psi^a
= (\Bar\psi^a_-~\Bar\psi^a_+)$, whence the quadratic term in the
action can be written (up to a total derivative) as
\begin{equation*}
\eL_F^{(2)} = - \omega_{ab} \Bar\psi^a_+ \Gamma_m D_m \psi^b_-~.
\end{equation*}
Finally the quartic terms in the action are given by
\begin{equation}\label{eq:LF4}
\eL_F^{(4)} = -\tfrac1{12} \Omega_{abcd} \left(
\Bar\psi^a_-\Gamma_m\psi^b_+\, \Bar\psi^c_-\Gamma_m\psi^d_+ -
\Bar\psi^a_+\psi^b_+\, \Bar\psi^c_-\psi^d_- \right)~.
\end{equation}

This action is invariant under the following supersymmetry
transformations:
\begin{align}\label{eq:4dsusy}
\delta_\varepsilon \phi^i & = \gamma^i_{Aa} \left( \Bar\varepsilon^A_+
\psi^a_+ + \Bar\varepsilon^A_- \psi^a_- \right)\notag\\
\delta_\varepsilon \psi^a_+ & = \gamma^{Aa}_i \d_m\phi^i \Gamma_m
\varepsilon_{-\,A} - \Hat\Gamma_i{}^a{}_b \delta_\varepsilon \phi^i
\psi^b_+\notag\\
\delta_\varepsilon \psi^a_- & = - \gamma^{Aa}_i \d_m\phi^i \Gamma_m
\varepsilon_{+\,A} - \Hat\Gamma_i{}^a{}_b \delta_\varepsilon \phi^i
\psi^b_-~.
\end{align}

\section{A cohomological theory for hyperinstantons}

In this section we define the dimensionally reduced theory not on
$\EE^4$ but on a four-dimensional spin manifold $M$.  As usual the theory
extends by covariantising the derivatives, but the supersymmetry
transformations will not be a symmetry unless the parameter is
covariantly constant; whence $M$ must admit parallel spinors.  In 4
dimensions, this means that $M$ is hyperk\"ahler or flat.  The flat
case is not interesting  for our present purposes since the holonomy
is trivial and we cannot guarantee that the supersymmetry will square
to zero (even on shell and up to `gauge' transformations: here
diffeomorphisms on $X$ or local $\Sp(n)$ transformations).  Therefore
we will assume from now on that $M$ is a four-dimensional hyperk\"ahler
manifold.  For example we could choose $X$ to be an ALE space or $K3$;
although the present analysis is general.  Reducing on such a manifold
we will arrive at the topological sigma model introduced by Anselmi
and Fr\`e \cite{AF}.

\subsection{The theory on a hyperk\"ahler manifold}

The holonomy of a generic four-dimensional riemannian manifold is
$\SO(4) \cong \Sp(1)\cdot\Sp(1)$.  For a hyperk\"ahler manifold, the
holonomy reduces to one of the $\Sp(1)$ factors.  If we identify the
Lie algebra $\so(4)$ with the adjoint representation $\bigwedge^2$,
then the split $\so(4) \cong \sp(1) \times \sp(1)$ corresponds to the
vector space decomposition $\bigwedge^2 = \bigwedge_+^2 \oplus
\bigwedge_-^2$ into self-dual and anti-self-dual two-forms.  We will
write $\SO(4) = \Sp(1)_+\cdot \Sp(1)_-$ to reflect this fact.  The
spin cover is a direct product $\Spin(4) = \Sp(1)_+ \times \Sp(1)_-$
which agrees with the notation in Section \ref{sec:4dspinors}.

For definiteness we choose the holonomy group to be $\Sp(1)_+$.  This
means that the negative chirality spinors are singlets of the holonomy
group and hence `scalars' in the new theory.  The surviving
supersymmetry will then be the one with $\varepsilon^A_-$ as
parameter.  We will let $\theta^A$ be a commuting parallel spinor on
$M$.  The index $A$ is a $\Sp(1)_T$ index which we carry simply to
remind ourselves that $\theta^A$ obeys a reality condition involving
$\epsilon^{AB}$. Notice that for commuting spinors $\Bar\theta^A
\theta^B = - \Bar\theta^B \theta^A$, whence
\begin{equation*}
\Bar\theta^A \theta^B = \tfrac12
\epsilon_{CD}\Bar\theta^C\theta^D\,\epsilon^{AB} = \tfrac12 \Bar\theta^C
\theta_C \epsilon^{AB}~.
\end{equation*}
We choose to normalise $\Bar\theta^A\theta_A=2$, whence
$\Bar\theta^A\theta^B = \epsilon^{AB}$.  In terms of this commuting
spinor, we can write the surviving supersymmetry parameter
$\varepsilon^A_- = \varepsilon \theta^A$, where $\varepsilon$ is an
anticommuting parameter.  Notice that $\theta^A$ has four real
components, whence we have four scalar supercharges.  The
supersymmetry transformations that we will write down are generated by
a particular linear combination, which if necessary may be broken down
into its constituents.  In fact, it is easy to see that the
independent supersymmetries are given by the combinations $Q_- =
\bar\theta^A Q_{A-}$ and $Q_-^{AB} = \bar\theta^{(A}Q^{B)}_-$; the
former being the scalar supersymmetry relevant to our deliberations
here.

In order to write the action and the BRST symmetry it will be
convenient to decompose our fields into irreducible representations of
the holonomy group.  To this end, the parallel spinor proves very
useful.  Indeed, let us define the following fields:
\begin{equation*}
\xi^{Aa} \equiv \Bar\theta^A \psi_-^a\qquad\text{and}\qquad
\chi^{Aa}_m \equiv \Bar\theta^A \Gamma_m\psi_+^a~.
\end{equation*}
In order to invert these definitions it will be necessary to use the
ubiquitous Fierz identity:
\begin{equation}\label{eq:fierz}
\theta^A \Bar\theta^B = -\tfrac14 \epsilon^{AB} \left(\1 -
\Gamma_5\right) - \tfrac18 \sK_{mn}^{AB}\Gamma_{mn}~,
\end{equation}
where we have defined
\begin{equation}\label{eq:jdefined}
\sK_{mn}^{AB} \equiv \Bar\theta^A\Gamma_{mn}\theta^B~.
\end{equation}
It obeys the following relations:
\begin{equation}\label{eq:jselfdual}
\sK_{mn}^{AB} = \sK_{mn}^{BA}\qquad\text{and}\qquad
\sK_{mn}^{AB} = \tfrac12 \epsilon_{mnpq} \sK_{pq}^{AB}~.
\end{equation}
In other words, $\sK_{mn}^{AB}$ transforms under the representation
$\bigwedge^2_+ \otimes \sigma^2$ of $\SO(4)\times\Sp(1)_T$, where
$\sigma^2 \equiv \bigodot^2\sigma^1$ is the adjoint representation.

The field $\chi^{Aa}_m$ defined above has more degrees of freedom than
$\psi^a_+$, whence we expect that it obeys a constraint.  In fact, it
does:
\begin{equation}\label{eq:chiconstraint}
\sK_{mn}{}^A{}_B \chi^{Ba}_n = - 3 \chi^{Aa}_m~,
\end{equation}
where we have used that $\theta_C\Bar\theta^C = \tfrac12 (\1 -
\Gamma_5)$, which follows from the Fierz identity \eqref{eq:fierz}.
Using this fact, we may invert the definition of $\chi^{Aa}_m$ to
deduce
\begin{equation*}
\psi^a_+ = - \tfrac14 \chi^{Aa}_m \Gamma_m \theta_A \qquad\implies
\Bar\psi^a_+ = \tfrac14 \chi^{Aa}_m \Bar\theta_A \Gamma_m~.
\end{equation*}
Similarly, contracting $\xi^{Aa}$ with $\theta_A$ we get
\begin{equation*}
\psi^a_- = \xi^{Aa}\theta_A\qquad\implies
\Bar\psi^a_- = \Bar\theta_A\xi^{Aa}~.
\end{equation*}

One might be tempted to define the field $B_{mn}^{Aa} \equiv
\Bar\theta^A \Gamma_{mn}\psi_-^a$, but in view of the last equation,
$B_{mn}^{Aa}$ would be simply re-expressed in terms of $\xi^{Aa}$:
$B_{mn}^{Aa} = \sK_{mn}{}^A{}_B \xi^{Ba}$.

\subsection{A cohomological theory}

We can now write the theory in terms of the new fields and check that
the remaining supersymmetry does indeed square to zero (at least
on-shell).  We first rewrite the action in terms of the new fields.

Clearly $\eL_B$ does not change, whereas the quadratic term
$\eL_F^{(2)}$ becomes
\begin{equation*}
\eL_F^{(2)} = \omega_{ab} \epsilon_{AB} \chi^{Aa}_m \sP_{mn}{}^B{}_C
D_n\xi^{Cb}~.
\end{equation*}
where $\sP$ is the projector onto those $\chi^{Aa}_m$ which obey
\eqref{eq:chiconstraint}.  This projector can be found as follows.
Using the Fierz identity \eqref{eq:fierz}, one can show that
\begin{equation}\label{eq:Jdef}
\sK_{mn}{}^A{}_B \sK_{np}{}^B{}_C = 3 \delta_{mp} \delta^A{}_C - 2
\sK_{mp}{}^A{}_C~,
\end{equation}
from where it follows that the map $\sK$ has eigenvalues $1$ and $-3$.
The projector onto the $-3$ eigenspace, to which $\chi^{Aa}_m$
belongs, is then given by
\begin{equation*}
\sP_{mn}{}^A{}_B = \tfrac14 \left(\delta_{mn}\delta^A{}_B -
\sK_{mn}{}^A{}_B \right)~.
\end{equation*}

The quartic terms are a little more complicated.  We record here the
following expressions
\begin{align*}
\Bar\psi^a_- \psi^b_- &= \epsilon_{AB} \xi^{Aa} \xi^{Bb}\\
\Bar\psi^a_+ \psi^b_+ &= \tfrac14 \epsilon_{AB}
\chi^{Aa}_m\chi^{Bb}_m\\
\Bar\psi^a_-\Gamma_m\psi^b_+ & = \epsilon_{AB} \xi^{Aa} \chi_m^{Bb}~,
\end{align*}
where we have once again used \eqref{eq:chiconstraint} in the last two
lines.  It is now a simple matter to plug these expressions into
\eqref{eq:LF4} to obtain
\begin{equation*}
\eL_F^{(4)} = \tfrac1{16} \epsilon_{AB}\epsilon_{CD}\,
\Omega_{abcd}\,\chi^{Aa}_m\chi^{Bb}_m \, \xi^{Cc} \xi^{Dd}~.
\end{equation*}

The action $\eL = \eL_B + \eL_F^{(2)} + \eL_F^{(4)}$ is invariant
under the following fermionic symmetry, obtained from
\eqref{eq:4dsusy} after putting $\varepsilon_+^A = 0$ and
$\varepsilon_-^A = \varepsilon \theta^A$:
\begin{align}\label{eq:hksusytrans}
\delta \phi^i & = \gamma^i_{Aa} \xi^{Aa} \equiv \xi^i\notag\\
\delta \chi^{Aa}_m & = \sP_{mn}{}^A{}_B \left( -4 \gamma^{Ba}_i
\d_n\phi^i - \Hat\Gamma_i{}^a{}_b \xi^i \chi^{Bb}_n \right)\notag\\
\delta \xi^{Aa} & = - \Hat\Gamma_i{}^a{}_b \xi^i \xi^{Ab}~.
\end{align}
Notice that we have dropped the explicit dependence of $\varepsilon$
and hence understand $\delta$ as a fermionic symmetry from now on.

Let us check $\delta^2$ on the fields.  Remembering that
$\gamma^i_{Aa}$ depends on $\phi^i$, we obtain
\begin{align*}
\delta^2 \phi^i &= \d_j\gamma^i_{Aa} \xi^j\xi^{Aa} - \gamma^i_{Aa} 
\Hat\Gamma_j{}^a{}_b \xi^j \xi^{Ab}\\
&= - \Gamma_j{}^i{}_k \xi^j\xi^k\tag{by \eqref{eq:gcc}}~,
\end{align*}
which vanishes because the Levi-Civita connection is torsionless.
Writing $\xi^{Aa}$ for $\gamma_i^{Aa}\xi^i$ and using that $\delta^2$
is a derivation we see that $\delta^2\xi^{Aa} =
(\delta^2\gamma_i^{Aa}) \xi^i + \gamma_i^{Aa}\delta^2\xi^i$; but this
vanishes since $\delta\xi^i = \delta^2\phi^i=0$.  Finally, we tackle
$\delta^2\chi^{Aa}_m$.  It will turn out that this is zero up to
equations of motion, which provides a double-check of the factor of
the quartic term in the action.  A little calculation yields
\begin{equation*}
\delta^2\chi^{Aa}_m = -4 \sP_{mn}{}^A{}_B D_n \xi^{Ba} - \tfrac12
\Hat R_{ij}{}^a{}_b \xi^i \xi^j \chi^{Ab}_m~.
\end{equation*}
On the other hand, under a variation of $\chi^{Aa}_m$, the action
varies like
\begin{equation*}
\delta\eL = \epsilon_{AB} \delta\chi^{Aa}_m \left( \omega_{ab}
\sP_{mn}{}^B{}_C D_n\xi^{Cb} + \tfrac18  \epsilon_{CD} \Omega_{abcd}
\chi^{Bb}_m \xi^{Cc} \xi^{Dd} \right)
\end{equation*}
In other words, putting $\chi^{Aa}_m$ on shell, means imposing the
equation of motion:
\begin{equation*}
\sP_{mn}{}^B{}_C D_n\xi^C{}_a = -\tfrac18 \epsilon_{CD}
\Omega_{abcd} \chi^{Bb}_m\xi^{Cc} \xi^{Dd}~;
\end{equation*}
whence the second variation of $\chi^{Aa}_m$ is given on shell by
\begin{equation*}
\delta^2\chi^{A}_{m\,a} =  \tfrac12 \epsilon_{CD}
\Omega_{abcd} \chi^{Ab}_m\xi^{Cc} \xi^{Dd}  - \tfrac12 \Hat
R_{ij}{}^a{}_b \xi^i \xi^j \chi^{Ab}_m~.
\end{equation*}
But this is zero by equation \eqref{eq:curvs}.

In order to make $\delta^2\chi^{Aa}_m=0$ on shell, it is necessary to
introduce a bosonic auxiliary field $H^{Aa}_m$ and define
\begin{align*}
\delta\chi^{Aa}_m &= H^{Aa}_m - \Hat\Gamma_i{}^a{}_b
\xi^i\chi^{Ab}_m\\
\delta H^{Aa}_m &= \tfrac12 \Hat R_{ij}{}^a{}_b \xi^i\xi^j\chi^{Ab}_m -
\Hat\Gamma_i{}^a{}_b \xi^i H^{Ab}_m~,
\end{align*}
where the last line is such that $\delta^2 \chi^{Aa}_m = 0$.  For this
transformation to make sense, $H^{Aa}_m$ must satisfy the analogous
constraint to \eqref{eq:chiconstraint}:
\begin{equation*}
\sK_{mn}{}^A{}_B H^{Ba}_n = -3 H^{Aa}_m~.
\end{equation*}
To show that $\delta^2 H^{Aa}_m = 0$ it suffices to express $H^{Aa}_m$
in terms of $\delta\chi^{Aa}_m$ and to use the fact that $\delta^2$ is
a derivation which has been shown to be zero already on $\chi^{Aa}_m$,
$\phi^i$ and $\xi^i$.  In summary, $\delta^2=0$ off shell.

This means that we should be able to write the action as a
`topological' term plus a term $\delta\Xi$, for some `gauge fermion'
$\Xi$, which because $\delta^2=0$ will be $\delta$-invariant.  Indeed,
consider the following expression
\begin{equation*}
\Xi = -\tfrac18 \epsilon_{AB} \omega_{ab} \chi^{Aa}_m \left( H^{Bb}_m
+ 8 \gamma^{Bb}_i \d_m\phi^i \right)~.
\end{equation*}
Its variation is given by
\begin{multline*}
\delta\Xi = -\tfrac18 \epsilon_{AB} \omega_{ab} H^{Aa}_m \left(
H^{Bb}_m + 8 \gamma^{Bb}_i \d_m\phi^i \right) + \epsilon_{AB}
\omega_{ab} \chi^{Aa}_m D_m\xi^{Bb} \\
+ \tfrac1{16} \epsilon_{AB} \epsilon_{CD} \Omega_{abcd} \chi^{Aa}_m
  \chi^{Bb}_m \xi^{Cc} \xi^{Dd}~,
\end{multline*}
where we have once again used \eqref{eq:curvs}.  Varying with respect
to $H^{Aa}_m$ we see that its equation of motion is algebraic, with
solution:
\begin{equation*}
H^{Aa}_m = -4 \sP_{mn}{}^A{}_B \gamma^{Ba}_i \d_n\phi^i~,
\end{equation*}
consistent with the supersymmetry variation of $\chi^{Aa}_m$ given in
\eqref{eq:hksusytrans}.  Plugging this back into $\delta\Xi$ we obtain
\begin{multline*}
\delta \Xi = 2 \omega_{ab} \sP_{mn\,AB} \gamma^{Aa}_i \gamma^{Bb}_j
\d_m\phi^i \d_n\phi^j\\
+ \epsilon_{AB} \omega_{ab} \chi^{Aa}_m D_m\xi^{Bb} + \tfrac1{16}
\epsilon_{AB}\epsilon_{CD}\, \Omega_{abcd}\,\chi^{Aa}_m\chi^{Bb}_m \,
\xi^{Cc} \xi^{Dd}~;
\end{multline*}
whence it follows that
\begin{equation*}
\eL = \tfrac12 \omega_{ab} \sK_{mn\,AB} \gamma^{Aa}_i \gamma^{Bb}_j
\d_m\phi^i \d_n\phi^j + \delta\Xi~.
\end{equation*}
This action was discovered by Anselmi and Fr\`e \cite{AF} by the more
traditional method of gauge fixing a topological symmetry.  Here we
have rediscovered it simply as the dimensional reduction of
the six-dimensional supersymmetric sigma model on a four-dimensional
hyperk\"ahler manifold.

\subsection{Topological interpretation}

In order to elucidate the topological meaning of this theory, it is
first of all convenient to redefine our fields slightly.  To this end
let us define
\begin{equation*}
\chi^i_m = \gamma^i_{Aa} \chi^{Aa}_m \qquad\text{and}\qquad
H^i_m = \gamma^i_{Aa} H^{Aa}_m~,
\end{equation*}
subject to the constraints
\begin{equation*}
h^{np} \sK_{mn}{}^i{}_j \chi^j_p = -3 \chi^i_m
\qquad\text{and}\qquad
h^{np} \sK_{mn}{}^i{}_j H^j_p = -3 H^i_m~,
\end{equation*}
where
\begin{equation}\label{eq:kdefined}
\sK_{mn}{}^i{}_j = \sK_{mn}{}^A{}_B \gamma^i_{Aa} \gamma_j^{Ba}~,
\end{equation}
and where we have introduced a metric $h_{mn}$ on $M$. In terms of
these fields, the action in curved space becomes
\begin{equation*}
\eS = \int_M d^4x \sqrt{h} h^{mn} \left( \tfrac12 g_{ij} \d_m\phi^i
\d_n\phi^j + g_{ij} \chi^i_m D_n\xi^j + \tfrac1{16}
R_{ijk\ell} \chi^i_m\chi^j_n \xi^k\xi^\ell \right)~.
\end{equation*}
This action is invariant under the following fermionic symmetry:
\begin{alignat*}{2}
\delta\phi^i &= \xi^i  & \qquad \delta\xi^i &= 0\\
\delta\chi^i_m &= H^i_m - \Gamma_j{}^i{}_k \xi^j \chi^k_m
& \qquad \delta H^i_m &= \tfrac12 R_{jk}{}^i{}_\ell \xi^j\xi^k
\chi^\ell_m - \Gamma_j{}^i{}_k \xi^j H^k_m~.
\end{alignat*}
Defining the integrated gauge fermion $\Theta$ by
\begin{equation*}
\Theta = - \tfrac18 \int_M d^4x \sqrt{h} g_{ij} h^{mn} \chi^i_m \left(
H^j_n + 8 \d_n\phi^j \right)~,
\end{equation*}
we have that after solving for the auxiliary field $H^i_m$, the action
is written as $\eS = \eS_{\mathrm{top}} + \delta\Theta$, where
\begin{equation*}
\eS_{\mathrm{top}} = \tfrac12 \int_M d^4x \sqrt{h} g_{ij} h^{mn} h^{pq}
\d_m\phi^i \sK_{np}{}^j{}_k \d_q\phi^k~.
\end{equation*}

Standard arguments now show that this cohomological theory localises
on the moduli space of maps $\phi:M\to X$ satisfying the following
condition:
\begin{equation}\label{eq:hyperinstanton}
h^{np} \sK_{mn}{}^i{}_j \d_p\phi^j = \d_m\phi^i~,
\end{equation}
namely that the projector $\sP$ annihilates $\d_m\phi^i$.  It is easy
to prove that maps $\phi:M\to X$ satisfying this condition minimise
the sigma model action:
\begin{equation*}
\eS[\phi] = \tfrac12 \| d\phi \|^2 \equiv \tfrac12 \int_M d^4x \sqrt{h}
h^{mn} g_{ij} \d_m \phi^i \d_n\phi^j~.
\end{equation*}
Indeed consider the inequality
\begin{equation*}
\|d\phi - \sK\,d\phi\|^2 \geq 0~.
\end{equation*}
Expanding and using equation \eqref{eq:Jdef} we find that
\begin{equation*}
\eS[\phi] \geq \eS_{\mathrm{top}}~,
\end{equation*}
with equality holding if and only if equation
\eqref{eq:hyperinstanton} is satisfied.  Such maps are called {\em
hyperinstantons\/} in \cite{AF}.  Because $\eS_{\mathrm{top}}$ is an
absolute minimum of the action, it should be a homotopy invariant.  We
will see that this is indeed the case. 

These maps are also called {\em triholomorphic\/} by analogy with the
case of two-dimensional topological sigma models.  In order to
understand this better it is necessary to look a little closer at the
map $\sK$.  As shown in Appendix \ref{sec:SO3}, $\sK_{mn}{}^i{}_j$ can
be written as
\begin{equation}\label{eq:explicitJ}
\sK_{mn\,ij} = - \sA_{\alpha\beta}\, \eI^\alpha_{mn} \eJ^\beta_{ij}~,
\end{equation}
where $\eI^\alpha_{mn}$ and $\eJ^\alpha_{ij}$ are the K\"ahler forms
corresponding to the complex structures in $M$ and $X$ respectively,
$\alpha$ and $\beta$ run from 1 to 3, and $\sA_{\alpha\beta}$ are the
entries of a matrix in $\SO(3)$.  The value of this matrix is not
significant, as it will change for a different choice of parallel
spinors.  We can explain its appearance as follows.  The choice of
complex structures on a hyperk\"ahler manifold is not canonical.  The
only object that has any invariant meaning is the two-sphere of
complex structures.  Any two choices of complex structures will be
related precisely by a matrix $\sA_{\alpha\beta}$ in $\SO(3)$.
Plugging the expression \eqref{eq:explicitJ} for $\sK$ into
\eqref{eq:hyperinstanton}, we find
\begin{equation}\label{eq:triholo}
- \sA_{\alpha\beta}\, (\eI^\alpha)_m{}^n (\eJ^\beta)^i{}_j \d_n\phi^j =
  \d_m\phi^i~,
\end{equation}
which has the following interpretation
\begin{equation}\label{eq:triholoabs}
\sA_{\alpha\beta}\, \eJ^\beta \circ \phi_* \circ \eI^\alpha = - \phi_*~,
\end{equation}
where $\phi_*: TM \to TX$ is the tangent map. This equation is a
quaternionic analogue of the Cauchy--Riemann equation defining
holomorphic maps.   Thus we way that a map $\phi:M\to X$ between
hyperk\"ahler manifolds is {\em triholomorphic} if its derivative
obeys \eqref{eq:triholoabs} for some matrix
$\sA_{\alpha\beta}\in\SO(3)$, and when $M$ is four-dimensional we call
its image $\phi(M) \subset X$ a {\em triholomorphic curve}.

It is possible to choose complex structures $\eI^\alpha$ for $M$ and
$\eJ^\alpha$ for $X$ in such a way that the matrix $\sA_{\alpha\beta}$
becomes the identity matrix.  In this case, the triholomorphicity
condition becomes
\begin{equation*}
\sum_\alpha \eJ^\alpha \circ \phi_* \circ \eI^\alpha = - \phi_*~.
\end{equation*}
One might be tempted to think that a more sensible notion of
triholomorphicity would demand that $\phi_*$ commutes with the action
of all three complex structures, just like for holomorphic maps; but
the nonexistence of a preferred point in the two-sphere of complex
structures means that there is no way to canonically identify the
two-spheres of complex structures in $M$ and $X$.  Even if we take $M$
and $X$ with a fixed hyperk\"ahler structure, the resulting equations
would form an overconstrained system, whereas equation
\eqref{eq:triholo} is elliptic.

As a final remark, let us show that $\eS_{\mathrm{top}}$ is indeed a
homotopy invariant.  From the explicit expression \eqref{eq:explicitJ}
for $\sK$, we can write $\eS_{\mathrm{top}}$ as follows:
\begin{equation*}
\eS_{\mathrm{top}} = - \tfrac12 \int_M d^4x \sqrt{h} h^{mn} h^{pq}
\sA_{\alpha\beta} \eI^\alpha_{np} \eJ^\beta_{ij} \d_m\phi^i
\d_q\phi^j~.
\end{equation*}
Using the self-duality of the K\"ahler forms $\eI^\alpha_{mn}$ on $M$,
we can write this more invariantly as
\begin{equation*}
\eS_{\mathrm{top}} = - \int_M \sA_{\alpha\beta}\, \eI^\alpha \wedge
\phi^*\eJ^\beta~,
\end{equation*}
which is clearly a homotopy invariant since the two-forms $\eI^\alpha$
and $\eJ^\alpha$ are closed.

\section{Octonionic instantons and triholomorphic curves}

In this section we specialise to the case where the target space $X$
is the moduli space $\eM_K$ of Yang--Mills instantons of a
four-dimensional hyperk\"ahler manifold $K$ and in this way we will
recover a gauge-theoretical result mentioned in \cite{DT,Thomas}.
This is the higher-dimensional analogue of the relation between
instantons and holomorphic curves in four-dimensional gauge theory
\cite{DS1,DS2,BJSV}.  This relation says the following.  Consider
Yang--Mills theory on a four-manifold $\Sigma_1 \times \Sigma_2$,
where $\Sigma_i$ are two-dimensional surfaces.  Rescaling the metric
of $\Sigma_2$, we find that in the limit of zero size, instantons on
$\Sigma_1 \times \Sigma_2$ are in one-to-one correspondence with
holomorphic curves $\phi(\Sigma_1)$ in the moduli space of flat
connections on $\Sigma_2$.  The higher-dimensional analogue of this
relation, which we will discuss here, concerns octonionic instantons
on the eight-dimensional manifold $M \times K$ where both $M$ and $K$
are four-dimensional hyperk\"ahler manifolds.  In the limit of
shrinking $K$, these instantons are in one-to-one correspondence with
triholomorphic curves $\phi(M)$ on the moduli space $\eM_K$ of
instantons.

\begin{figure}[h]
\centering
\setlength{\unitlength}{1pt}
\begin{picture}(360,270)(0,0)
\put(155,250){\framebox(50,20){$\yms$}}
\put(0,210){\framebox(120,20){$\yms$ \quad $\MM^{5{+}1} \times
\varepsilon K$}}
\put(50,175){\makebox(60,20){$\lim\limits_{\varepsilon\to0}$}}
\put(0,140){\framebox(120,20){$\mathsf{S\sigma m}_6\quad\MM^{5{+}1} \to
\eM_K$}}
\put(0,80){\framebox(120,20){$\mathsf{S\sigma m}_4\quad M \to \eM_K$}}
\put(60,20){\oval(120,40){}}
\put(0,20){\makebox(120,20){triholomorphic}}
\put(0,0){\makebox(120,20){maps $M\to\eM_K$}}
\put(240,160){\framebox(120,40){}}
\put(240,180){\makebox(120,20){$\mathsf{SYM}_8$ \quad $Y$}}
\put(240,160){\makebox(120,20){\small
$\text{Hol}(Y)\subset\Spin(7)$~}}
\put(240,80){\framebox(120,20){$\mathsf{SYM}_8$ \quad $M \times
\varepsilon K$}}
\put(300,20){\oval(120,40){}}
\put(240,20){\makebox(120,20){octonionic}}
\put(240,00){\makebox(120,20){instantons}}
\put(155,40){\makebox(50,20){$\lim\limits_{\varepsilon\to0}$}}
\thicklines
\put(150,260){\line(-1,0){90}}
\put(60,260){\vector(0,-1){25}}
\put(60,205){\vector(0,-1){40}}
\put(60,135){\vector(0,-1){30}}
\put(0,50){\makebox(120,20){localising on}}
\put(210,260){\line(1,0){90}}
\put(300,260){\vector(0,-1){55}}
\put(300,155){\vector(0,-1){50}}
\put(240,50){\makebox(120,20){localising on}}
\put(230,30){\vector(-1,0){100}}
\put(130,10){\vector(1,0){100}}
\thinlines
\end{picture}
\caption{The correspondence between octonionic instantons and
triholomorphic curves starting from ten-dimensional supersymmetric
Yang--Mills.\label{fig:diagram}}
\end{figure}
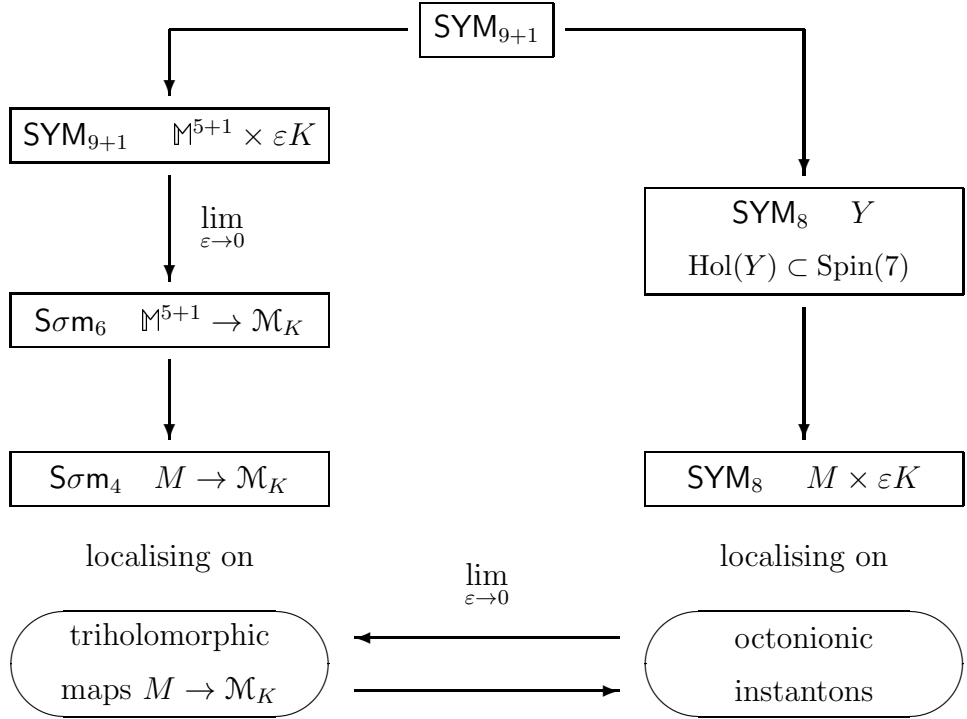

The way we derive this result is summarised in
Figure~\ref{fig:diagram}.  The starting point is ten-dimensional
supersymmetric Yang--Mills theory ($\yms$).  On the one hand,
corresponding to the left-hand side of the Figure, we can
dimensionally reduce it to six dimensions on a small
four-di\-men\-sion\-al hyperk\"ahler manifold $K$.  The adiabatic
limit of this theory is the supersymmetric sigma model on the
instanton moduli space $\eM_K$, which is well-known to possess a
hyperk\"ahler metric.  This means that we can apply the results from
the previous section.  As shown above, the six-dimensional
supersymmetric sigma model ($\mathsf{S\sigma m}$) with target $\eM_K$,
dimensionally reduced on a four-dimensional hyperk\"ahler manifold $M$
is a topological sigma model ($\mathsf{T\sigma m}$) localising on the
moduli space of triholomorphic maps $M \to \eM_K$, or hyperinstantons.
On the other hand, corresponding to the right-hand side of the Figure,
we can reduce $\yms$ directly on the eight-dimensional manifold
$M\times K$ of holonomy $\Sp(1)\times\Sp(1) \subset \Spin(7)$.  As
shown in \cite{AFOS,BTESYM} the resulting cohomological theory
localises on the moduli space of octonionic instantons.  Taking the
adiabatic limit in which $K$ shrinks to zero size, our argument shows
that we recover the topological sigma model $M \to \eM_K$.  We will
also see this directly.

\subsection{Supersymmetric Yang--Mills and hyperinstantons}
\label{sec:symhyper}

Let us start with $\yms$ and reduce it on $\MM^{5{+}1} \times K$,
where $K$ is a four-di\-men\-sion\-al hyperk\"ahler manifold.  If $K$
is not flat, then it preserves only one half of the supersymmetry.  In
the limit in which $K$ shrinks to zero size, the resulting effective
theory is a six-dimensional theory with eight real supercharges.  We
will argue that this is a supersymmetric sigma model with target space
the instanton moduli space $\eM_K$ on $K$.  Parenthetically, this
shows that $\eM_K$ admits a hyperk\"ahler metric, a fact that is
well-known and can of course be proven without recourse to
supersymmetry.  We will actually show that the bosonic sector of the
effective six-dimensional theory is a sigma model on the instanton
moduli space.  Together with the fact that the effective theory is
supersymmetric, this means that the only natural candidate is the
supersymmetric sigma model.  A fuller treatment is possible but falls
outside the scope of the present paper.

Let us consider the bosonic sector of $\yms$ on $\MM^{5{+}1} \times
K$, which is described by the Yang--Mills action.  In differential
form notation we will refer to the components of the gauge field along
$\MM\equiv \MM^{5{+}1}$ and $K$ as $A_{\MM}$ and $A_K$ respectively.
They take values in a Lie algebra $\fg$ with a fixed invariant metric
denoted by $\Tr$.  The Yang--Mills action on $\MM \times K$ can be
written as
\begin{equation*}
\eLYM = \tfrac14 \|F_K\|^2 + \tfrac12 \|d_{\MM} A_K - D_K A_{\MM}\|^2
+ \tfrac14 \|F_{\MM}\|^2~,
\end{equation*}
where $F_K$ and $F_{\MM}$ are the components of the field-strength
along the $K$ and $\MM$ directions respectively, $d$ and $D$ stand
for the exterior and exterior covariant derivatives respectively, and
$\|~\|^2$ is the pointwise norm on forms including also the metric on
the Lie algebra, e.g.,
\begin{equation*}
\|F_K\|^2 = \Tr F_K\wedge \star F_K~,
\end{equation*}
with $\star$ being the Hodge operator in $\MM \times K$.  We now take
the limit in which $K$ shrinks to zero size; that is, we rescale the
metric on $K$ by a parameter $\varepsilon$.  Performing this rescaling
we see that $\eLYM$ scales as
\begin{equation}\label{eq:scaledlag}
\eLYM^{(\varepsilon)} = \tfrac14 \|F_K\|^2 + \tfrac12 \varepsilon
\|d_{\MM} A_K - D_K A_{\MM}\|^2 + \tfrac14 \varepsilon^2
\|F_{\MM}\|^2~.
\end{equation}

Setting $\varepsilon = 0$ only the first term survives.  This term has
no derivatives along $\MM$ whence it is to be understood as an
effective potential.  The minima of this potential are precisely the
instantons on $K$.  The effective theory on $\MM$ induced by
$\eLYM^{(\varepsilon=0)}$ has therefore no dynamics.  In order to
obtain a nontrivial effective theory we must consider the next term in
the perturbative $\varepsilon$ expansion.  In this term, the field
$A_{\MM}$ has no derivatives along $\MM$, and hence it plays the
r\^ole of an auxiliary field.  Its only effect is to cancel the
components of $d_\MM A_K$ which lie in the image of $D_K$---that is,
which are tangent to the orbits of the gauge group.  In other words,
this leaves a theory of maps from $\MM$ to the instanton moduli space
$\eM_K$ on $K$.  We will now show that this theory is a sigma model.
The following discussion assumes familiarity with the local geometry
of the instanton moduli space $\eM_K$, which is summarised in Appendix
\ref{sec:geometry}.

We will let $x^\mu$ for $\mu=0,1,\ldots,5$ denote the coordinates on
$\MM$ and $y^m$ for $m=1,2,3,4$ denote the coordinates on $K$.  Let
$A_\mu(x,y)$ and $A_m(x,y)$ denote the different components of the
gauge fields.  Minimising the action at $\epsilon=0$ means that $F_K$
is self-dual or anti-self-dual.  Let us take for definiteness the case
that $F_K$ is self-dual.  For fixed $x$, $A_m(x,y)$ is a self-dual
gauge field, so it defines a map from $\MM$ to the space of
self-dual connections $\eA_+$.  In turn, a self-dual gauge field $A_m$
defines a map $\phi{=}[A]:\MM \to \eM_K$ defined by the composition
$\MM \to \eA_+ \to \eM_K$, where the second map is the natural
projection sending a connection to its gauge equivalence class and the
first map is the one induced by $A_m$.  Then we can break $\d_\mu A_m$
according to \eqref{eq:split} as follows:
\begin{equation}\label{eq:breakup}
\d_\mu A_m(x,y) = \d_\mu \phi^A(x) \delta_A A_m + D_m \epsilon_\mu~,
\end{equation}
where $\delta_A A_i$ are the components tangent to $\eM_K$ and
$\epsilon_\mu$ are some gauge parameters which are given in terms of
$\d_\mu A_m$ by solving
\begin{equation*}
D_m D_m \epsilon_\mu = D_m \d_\mu A_m~.
\end{equation*}

We now insert \eqref{eq:breakup} into the second term in
\eqref{eq:scaledlag} and we obtain
\begin{multline*}
\tfrac12 \int_\MM d^6x \, \langle \delta_A A,\delta_B A\rangle\, \d_\mu
\phi^A \d^\mu \phi^B\\
+ \tfrac12  \int_\MM d^6x \int_K d^4 y \sqrt{g_K} \Tr \left(D_m
  (\epsilon_\mu - A_\mu)\right)^2~.
\end{multline*}
The second term can be discarded after shifting $A_\mu$ and
integrating it out.  The first term is the action of a sigma model of
maps $\phi: \MM \to \eM_K$, with metric given by
\eqref{eq:instantonmetric}; that is,
\begin{equation*}
\eS^{\mathrm{eff}}_{\mathrm{bosonic}} = \tfrac12 \int_\MM d^6x \,
G_{AB}(\phi) \d_\mu \phi^A \d^\mu \phi^B~.
\end{equation*}
In other words the bosonic term of the effective action is a sigma
model with target the moduli space of instantons $\eM_K$ on $K$.
The fermionic terms make up a supersymmetric theory with 8 real
supercharges.  It can be argued that the unique such theory whose
bosonic term is the sigma model is the supersymmetric sigma model on
$\eM_K$.  Incidentally, this proves that the metric $G_{AB}(\phi)$
given by \eqref{eq:instantonmetric} is hyperk\"ahler.

We are now in a position to apply the results described in the
previous sections and reduce the theory on a four-dimensional
hyperk\"ahler manifold $M$.  The resulting cohomological field theory
is the topological sigma model of \cite{AF} which localises on the
space of triholomorphic maps $M \to \eM_K$; that is, on the
hyperinstantons.

\subsection{Supersymmetric Yang--Mills and octonionic instantons}

On the other hand we can reduce $\yms$ directly on the manifold
$M \times K$ and then take the adiabatic limit.  This
eight-dimensional manifold has holonomy $\Sp(1)\times \Sp(1) \subset
\Spin(7)$; hence by the results of \cite{AFOS} (see also
\cite{BTESYM}), the reduction of $\yms$ to $M\times K$ is a
cohomological theory which localises on the moduli space of octonionic
instantons.  This cohomological theory was originally discussed in
\cite{BKS,BKS2,AOS} from a different perspective.

On an eight-manifold $Y$ of $\Spin(7)$ holonomy, there exists a
self-dual parallel four-form $\Omega$.  Relative to a local $\Spin(7)$
coframe $\{\omega^a\}$, $\Omega$ is given by
\begin{multline}\label{eq:8d4form}
\Omega = \omega^{1234} + \omega^{1258} - \omega^{1267} + \omega^{1357}
+ \omega^{1368} - \omega^{1456} + \omega^{1478} \notag\\
+ \omega^{2356} - \omega^{2378} + \omega^{2457} + \omega^{2468} -
  \omega^{3458} + \omega^{3467} + \omega^{5678}~,
\end{multline}
where we have used the shorthand $\omega^{abcd} = \omega^a
\wedge \omega^b \wedge \omega^c \wedge \omega^d$.  This four-form is
intimately linked with the octonion algebra (see, e.g., \cite{HL});
but for our present purposes its most interesting property is that it
can be used to define an endomorphism of the two-forms:
\begin{equation*}
F \mapsto \star (\Omega \wedge F)~,\quad\text{or in components}\quad
F_{ab} \mapsto \tfrac12 \Omega_{abcd} F_{cd}~.
\end{equation*}
The octonionic instanton equations are given by the following linear
equations on the curvature \cite{CDFN,Ward},
\begin{equation}\label{eq:8instanton}
F_{ab} = \tfrac12 \Omega_{abcd} F_{cd}~.
\end{equation}
Relative to the local frame above, these equations expand to
\begin{align}\label{eq:8instantonfull}
F_{12} - F_{34} - F_{58} + F_{67} &= 0\notag\\
F_{13} + F_{24} - F_{57} + F_{68} &= 0\notag\\
F_{14} - F_{23} + F_{56} - F_{78} &= 0\notag\\
F_{15} + F_{28} + F_{37} - F_{46} &= 0\\
F_{16} - F_{27} + F_{38} + F_{45} &= 0\notag\\
F_{17} + F_{26} - F_{35} + F_{48} &= 0\notag\\
F_{18} - F_{25} - F_{36} - F_{47} &= 0~.\notag
\end{align}
It is possible to show that these equations are the zero locus of an
octonionic moment map defined by the action of the gauge group on the
space of gauge fields and thus that the moduli space of octonionic
instantons can be exhibited as an infinite-dimensional octonionic
K\"ahler quotient \cite{Moduli}, at least when the eight-dimensional
manifold is flat.

We now consider the manifold $M\times K$.  Each $M$ and $K$ are
hyperk\"ahler four-dimensional manifolds, hence they have each a
triplet of parallel two-forms corresponding to the K\"ahler forms of the
hyperk\"ahler structure.  Depending on the orientation, they can be
chosen to be self-dual or anti-self-dual.  We will fix the
orientations of $M$ and $K$ in such a way that the parallel forms are
anti-self-dual.  Let us choose the local $\Spin(7)$ frame above such
that $\omega^i$ for $i=1,2,3,4$ are a local $\Sp(1)$ frame for $M$
and $\omega^m$ for $m=5,6,7,8$ are a local $\Sp(1)$ frame for $K$.
Relative to this frame we can choose the anti-self-dual forms on $M$
and $K$ to be given respectively by the following expressions:
\begin{xalignat}{2}
\eI^1 &= \omega^{12} - \omega^{34} & \qquad \eIt^1 &= \omega^{56} -
\omega^{78}\notag\\
\eI^2 &= \omega^{13} + \omega^{24} & \qquad \eIt^2 &= \omega^{57} +
\omega^{68}\label{eq:choiceIJ}\\
\eI^3 &= \omega^{13} - \omega^{23} & \qquad \eIt^3 &= \omega^{58} -
\omega^{67}~.\notag
\end{xalignat}
In terms of these forms, the $\Spin(7)$-invariant four-form $\Omega$ is
given by
\begin{equation}\label{eq:omegaIJ}
\Omega = -\tfrac16 \eI^\alpha \wedge \eI^\alpha - \tfrac16 \eIt^\alpha
\wedge \eIt^\alpha + \sA_{\alpha\beta}\, \eI^\alpha \wedge \eIt^\beta~,
\end{equation}
where the matrix $\sA_{\alpha\beta}$ is given by $\sA = \left(
\begin{smallmatrix} 0 & 0 & 1\\ 0 & 1 & 0 \\ -1 & 0 & 0
\end{smallmatrix}\right)\in \SO(3)$.  Any other choice of $\eI^\alpha$
or $\eIt^\alpha$, consistent with the fact that they are a
hyperk\"ahler structure, is obtained from the one in
\eqref{eq:choiceIJ} by an $\SO(3)$ rotation.  The expression for
$\Omega$ in terms of the new $\eI^\alpha$ and $\eIt^\alpha$ would then
be given by \eqref{eq:omegaIJ} but where the matrix $A$ goes over to
$LAR^t$, where $L$ and $R$ are matrices in $\SO(3)$.  In other words,
$A$ becomes an arbitrary $\SO(3)$ matrix itself.

In Appendix \ref{sec:omega} (see equation \eqref{eq:Fij}) it is shown
that the first three instanton equations in \eqref{eq:8instantonfull}
can be rewritten as
\begin{equation*}
F \cdot \eI^\alpha = \sA_{\alpha\beta}\, F \cdot \eIt^\beta~,
\end{equation*}
where $F \cdot \eI^\alpha \equiv F_{ij} \eI^\alpha_{ij}$ and
similarly for $F \cdot \eIt^\alpha$.  Under a rescaling of the metric
in $K$ by $\varepsilon$, these equations scale as
\begin{equation}\label{eq:rescaledFij}
\varepsilon\, F \cdot \eI^\alpha = \sA_{\alpha\beta}\, F \cdot
\eIt^\beta~,
\end{equation}
whence the limit $\varepsilon \to 0$ implies that $F \cdot \eIt^\alpha
= 0$ for all $\alpha$.  In other words, in the limit of shrinking $K$,
for each point $x\in M$, the components of $F$ along $K$ are self-dual
and hence define an instanton.  Therefore the map $x\mapsto A_K(x,y)$
induces a map $\phi{=}[A_K]: M \to \eM_K$, which the remaining
octonionic instanton equations will constrain as follows.

As shown in Appendix \ref{sec:omega} (see equation \eqref{eq:Fim}) the
last four instanton equations in \eqref{eq:8instantonfull} can be
written as follows:
\begin{equation}\label{eq:remaining}
F_{im} = \sA_{\alpha\beta}\, \eI^\alpha_{ij} F_{jn} \eIt^\beta_{nm}~.
\end{equation}
In order to be able to interpret these equations in terms of the map
$\phi: M \to \eM_K$, we first expand the field strength:
\begin{equation*}
F_{im} = \d_i A_m - D_m A_i~.
\end{equation*}
Plugging this back into \eqref{eq:remaining} we obtain
\begin{equation*}
\d_i A_m - D_m A_i = \sA_{\alpha\beta} \eI^\alpha_{ij}
\left( \d_j A_n - D_n A_j \right) \eIt^\beta_{nm}~.
\end{equation*}
We now break up $\d_i A_m$ as in \eqref{eq:breakup} and rewrite this
equation as
\begin{multline*}
\d_i \phi^A \delta_A A_m - A_{\alpha\beta} \eI^\alpha_{ij}
\d_j \phi^A \delta_A A_n \eIt^\beta_{nm}\\
 = D_m \left( A_i - \epsilon_i \right) - \sA_{\alpha\beta}
 \eI^\alpha_{ij} \eIt^\beta_{nm} D_n \left( A_j - \epsilon_j
 \right)~.
\end{multline*}
We now contract with $\delta_B A_m$ and integrate over $K$ to obtain
\begin{equation}\label{eq:hyperinstantonredux}
\d_i\phi^A G_{AB} +  \sA_{\alpha\beta} \eI^\alpha_{ij} \d_j\phi^A
\eJ^\beta_{AB} = 0~,
\end{equation}
where we have used the expressions \eqref{eq:instantonmetric} for the
metric $G_{AB}$ and \eqref{eq:kaehlerM} for the complex structure
$\eJ^\beta_{AB}$ on $\eM$, as well as the identities
\eqref{eq:linsdalt} and \eqref{eq:gausslaw}.  Finally we can multiply
both sides of equation \eqref{eq:hyperinstantonredux} with the inverse
metric to obtain
\begin{equation*}
\d_i\phi^A  = - \sA_{\alpha\beta}\, \eI^\alpha_{ij}\, \d_j\phi^B\,
{\eJ^\beta}_B{}^A~,
\end{equation*}
which is precisely the triholomorphicity condition \eqref{eq:triholo}
on the map $\phi: M \to \eM_K$ making it a hyperinstanton.

Conversely, a triholomorphic map $\phi: M \to \eM_K$ lifts to a map
$A_m: M \to \eA_+$.  The variations $\d_i A_m$ do not necessarily obey
Gauss's law \eqref{eq:gausslaw}, but there exist gauge parameters
$\epsilon_i$ such that $\d_i A_m - D_m \epsilon_i$ do.  It is easy to
see that if $\phi$ is triholomorphic, then the pair $(\epsilon_i,
A_m)$, understood as a connection on $M\times K$, obeys the second set
of octonionic instanton equations \eqref{eq:Fim}.  The first set of
equations \eqref{eq:Fij} is not obeyed because $A_m$ are self-dual
connections; but as we saw, under a conformal rescaling of $K$, this
is a limit of solutions of \eqref{eq:Fij}.  Therefore one can argue
that in a neighbourhood of $(\epsilon_i,A_m)$ there is a solution of
the rescaled equations \eqref{eq:rescaledFij} at least for $\epsilon$
small enough.

In other words, we have shown that there is a one-to-one
correspondence between triholomorphic maps $M \to \eM_K$
(hyperinstantons) and octonionic instantons on $M \times K$ in the
limit when $K$ shrinks to zero size.  The correspondence follows from
an equivalence of cohomological theories obtained from $\yms$ by
dimensional reduction.

\section*{Acknowledgement}

It is a pleasure to thank Bobby Acharya, Jerome Gauntlett, Chris Hull,
Yolanda Lozano, Jim McCarthy, Sonia Stanciu, Arkady Vaintrob and
Nelson Vanegas for useful conversations, and Richard Thomas for some
comments on a previous version of the paper.  In addition we are
grateful to the EPSRC for support and {\sf bar \bfseries med} for
providing such a stimulating atmosphere.
 
\appendix

\section{Supersymmetry of the sigma model}\label{sec:ssm}

In this appendix we derive the factors in front of the fermionic terms
in the lagrangian \eqref{eq:ssm} for the supersymmetric sigma model.
The action \eqref{eq:ssm} is determined essentially by supersymmetry
alone.  Let us write the lagrangian $\eL$ as a sum of three terms
\begin{equation*}
\eL = \eL^{(0)} + \eL^{(2)} + \eL^{(4)}~,
\end{equation*}
where
\begin{align*}
\eL^{(0)} &= \tfrac{1}{2} g_{ij} \d_M \phi^i \d^M \phi^j\\
\eL^{(2)} &= \alpha \omega_{ab} \Bar\Psi^a \Hat\Gamma^M D_M \Psi^b\\
\eL^{(4)} &= \beta \Omega_{abcd} (\Bar\Psi^a \Hat\Gamma_M \Psi^b)
(\Bar\Psi^c \Hat\Gamma^M\Psi^d)~,
\end{align*}
and $\alpha$ and $\beta$ are parameters to be determined.  The
supersymmetry transformations \eqref{eq:6dsusy} contain terms of
orders $-1$ and $1$ in the fermions: $\delta_\varepsilon =
\delta^{(-1)} + \delta^{(1)}$.  Invariance under supersymmetry means
that up to a total derivative the following cancellations must take
place:
\begin{align*}
\delta^{(1)} \eL^{(0)} + \delta^{(-1)} \eL^{(2)} & =  0\\
\delta^{(1)} \eL^{(2)} + \delta^{(-1)} \eL^{(4)} & =  0\\
\delta^{(1)} \eL^{(4)}  & =  0~.
\end{align*}
The first and last identities do not require any special tricks: the
first one simply fixes $\alpha=\tfrac12$; whereas the last one is true
for any $\beta$ by virtue of the Bianchi identity for the curvature
tensor.  The middle identity is the crucial one, as it will fix
$\beta$.  After some calculation, one finds
\begin{equation*}
\delta^{(1)} \eL^{(2)} = \tfrac12 \Hat R_{ij\,ac} \d_M\phi^j
\gamma^i_{Ad} \Bar\varepsilon^A\Psi^d\,\Bar\Psi^a\Hat\Gamma^M\Psi^c~,
\end{equation*}
which after using \eqref{eq:curvs} becomes
\begin{equation*}
\delta^{(1)} \eL^{(2)} = \tfrac12 \Omega_{abcd} \d_M\phi^i
\gamma_i^{Aa} \Bar\Psi^b\Hat\Gamma^M\Psi^c\,\Bar\Psi^d\varepsilon_A~.
\end{equation*}
On the other hand, we have that
\begin{equation*}
\delta^{(-1)} \eL^{(4)} = 4 \beta \Omega_{abcd} \d_M\phi^i
\gamma_i^{Aa} \Bar\Psi^b\Hat\Gamma_N \Psi^c\, \Bar\Psi^d \Hat\Gamma^N
\Hat\Gamma^M \varepsilon_A~.
\end{equation*}
Therefore we are a Fierz away from being able to compare them.  The
required Fierz identity is the following:
\begin{equation*}
\Psi^a\Bar\Psi^b = \tfrac18 \Bar\Psi^b
\Hat\Gamma_M\Psi^a\,\Hat\Gamma^M (\1 - \Hat\Gamma_7) - \tfrac18
\tfrac1{3!} \Bar\Psi^b \Hat\Gamma_{MNP} \Psi^a \Hat\Gamma^{MNP}~.
\end{equation*}
Notice the important fact that the first term in the right hand side
is symmetric in $a$ and $b$, whereas the second term is
antisymmetric.  Using this fact we find that 
\begin{equation*}
\delta^{(1)} \eL^{(2)} = \tfrac1{12} \Omega_{abcd} \d_M\phi^i
\gamma_i^{Aa} \Bar\Psi^b\Hat\Gamma^N\Psi^c\,\Bar\Psi^d \Hat\Gamma^N
\Hat\Gamma^M \varepsilon_A~.
\end{equation*}
In other words, this will cancel the contribution from $\eL^{(4)}$
provided that $\beta = -\tfrac1{48}$ as in \eqref{eq:ssm}.  Notice
that this corrects a `typo' in an earlier version of \cite{FKS}.

\section{Expression for $\sK_{mn\,ij}$}\label{sec:SO3}

The object $\sK_{mn\,ij}$ was defined in \eqref{eq:kdefined} and obeys
relations coming from \eqref{eq:jselfdual}.  The second equation in
\eqref{eq:jselfdual} says that it is self-dual on its two $M$ indices,
whereas the first equation says that it is self-dual on its two $X$
indices.  From this fact it follows that the most general form for
$\sK$ is given by
\begin{equation*}
\sK_{mn\,ij} = - \sA_{\alpha\beta} \eI^\alpha_{mn} \eIt^\beta_{ij}~,
\end{equation*}
for some coefficients $\sA_{\alpha\beta}$.  From the fact that the
complex structures satisfy
\begin{equation*}
\eI^\alpha \circ \eI^\beta = \epsilon^{\alpha\beta\gamma} \eI^\gamma -
\delta^{\alpha\beta} \1\qquad\text{and}\qquad
\eIt^\alpha \circ \eIt^\beta = \epsilon^{\alpha\beta\gamma}
\eIt^\gamma - \delta^{\alpha\beta} \1~,
\end{equation*}
we see that equation \eqref{eq:Jdef} implies the following equations
on $\sA_{\alpha\beta}$:
\begin{equation}\label{eq:SO3}
\sA_{\alpha\beta} \sA_{\alpha\beta} = 3 \qquad\text{and}\qquad
\epsilon_{\alpha\gamma\eta} \epsilon_{\beta\delta\theta}
\sA_{\alpha\beta} \sA_{\gamma\delta} = 2 \sA_{\eta\theta}~.
\end{equation}
It is easy to see that these equations are solved by $\sA_{\alpha\beta}
= \delta_{\alpha\beta}$ and that the tangent space to the solution
space at that point is three-dimensional and given by antisymmetric
matrices.  Moreover it is also easy to show that if $\sA_{\alpha\beta}$
and $\sB_{\alpha\beta}$ are solutions then so is their product
$\sA_{\alpha\beta} \sB_{\beta\gamma}$.  It follows therefore that
elements of $\SO(3)$ solve these equations.  As we now show these are
the only solutions.  We will offer two proofs: one short and slick,
and one elementary.

\begin{lem}
A $3\times 3$ matrix $\sA_{\alpha\beta}$ solves equation \eqref{eq:SO3}
if and only if it is special orthogonal.
\end{lem}

\begin{proof}[Slick proof.]
Let $e_\alpha$ be a basis for $\RR^3$ turned into a Lie algebra
isomorphic to $\so(3)$ under the cross product.  Given
$\sA_{\alpha\beta}$ define a map $A: \RR^3 \to \RR^3$ by $A(e_\alpha)
= \sA_{\beta\alpha} e_\beta$, which by the second equation in
\eqref{eq:SO3} is a Lie algebra homomorphism.  Because $\so(3)$ is
simple, $A$ is either an automorphism or the zero map. The first
equation in \eqref{eq:SO3} discards this latter possibility. Because
$\so(3)$ is simple, its automorphism group is the adjoint group
$\SO(3)$.  Finally, any $\sA_{\alpha\beta}\in\SO(3)$ satisfies the
first equation in \eqref{eq:SO3}.
\end{proof}

\begin{proof}[Elementary proof.]
Expanding the $\epsilon$ symbols in the second equation in
\eqref{eq:SO3} we get the following equation
\begin{equation*}
\sA^2 - \sigma_1 \sA + \tfrac12 (\sigma_1^2 - \sigma_2) = \sA^t~,
\end{equation*}
where $\sigma_i = \Tr \sA^i$ and $\sA = [\sA_{\alpha\beta}]$ is a $3\times
3$ matrix.  Multiplying both sides of this equation with $\sA$ either on
the left or on the right and comparing with with the characteristic
equation for a $3\times 3$ matrix,
\begin{equation*}
\sA^3 - \sigma_1 \sA^2 + \tfrac12 (\sigma_1^2 - \sigma_2) \sA - (\Det \sA)\,\1
= 0~,
\end{equation*}
we see that $\sA^t\sA = \sA\sA^t = (\Det \sA)\,\1$.  Taking the trace and using
the first equation in \eqref{eq:SO3}, we see that $\Det \sA = 1$ and
that $\sA^t\sA = \sA\sA^t = \1$; whence $\sA\in \SO(3)$.
\end{proof}

\section{The octonionic instanton equations on $M\times K$}
\label{sec:omega}

In this appendix we rewrite the $\Spin(7)$ instanton equation
\eqref{eq:8instanton} for the explicit $\Omega$ built out of the
hyperk\"ahler structures of $M$ and $K$ and given by equation
\eqref{eq:omegaIJ}.  The octonionic instanton equation
\eqref{eq:8instanton} can be rewritten more invariantly as
\begin{equation}\label{eq:octinst}
F = \star (\Omega \wedge F)~,
\end{equation}
and it is this equation that we will work with.  We start by writing
all forms into components relative to the the $\Sp(1)$ structures of
$M$ and $K$.  We will let $i,j,k,l$ running from $1$ to $4$ be
``flat'' indices on $M$.  Similarly $m,n,p,q$ running from $5$ to $8$
are ``flat'' indices of $K$.  Indices $a,b,c,d$ are ``flat'' indices
of $M\times K$ and run from $1$ to $8$.  For the hyperk\"ahler
structures $\eI^\alpha$ and $\eIt^\alpha$ we have
\begin{equation*}
\eI^\alpha = \tfrac12 \eI^\alpha_{ab} \omega^{ab} = \tfrac12
\eI^\alpha_{ij} \omega^{ij}\qquad\text{and}\qquad
\eIt^\alpha = \tfrac12 \eIt^\alpha_{ab} \omega^{ab} = \tfrac12
\eIt^\alpha_{mn} \omega^{mn}~.
\end{equation*}
Similarly for the field strength we have $F = \tfrac12 F_{ab}
\omega^{ab}$.  In terms of these components and taking into account
\eqref{eq:omegaIJ}, the octonionic instanton equation
\eqref{eq:octinst} becomes
\begin{multline}\label{eq:8instantonIJ}
F_{ab} = - \tfrac1{48} \epsilon_{abcdefgh} \eI^\alpha_{cd}
\eI^\alpha_{ef} F_{gh} 
- \tfrac1{48} \epsilon_{abcdefgh} \eIt^\alpha_{cd}
\eIt^\alpha_{ef} F_{gh}\\
+ \tfrac18 \epsilon_{abcdefgh} \sA_{\alpha\beta} \eI^\alpha_{cd}
  \eIt^\beta_{ef} F_{gh}~.
\end{multline}

We now specialise the indices. Letting $(ab) = (ij)$ we find
\begin{equation*}
F_{ij} = - \tfrac1{48} \epsilon_{ijmnpqkl} \eIt^\alpha_{mn}
\eIt^\alpha_{pq} F_{kl} + \tfrac18 \epsilon_{ijklmnpq} \sA_{\alpha\beta}
\eI^\alpha_{kl} \eIt^\beta_{mn} F_{pq}~.
\end{equation*}
Using $\epsilon_{ijklmnpq} = \epsilon_{ijkl}\epsilon_{mnpq}$ and the
anti-self-duality of $\eI^\alpha$ and $\eIt^\alpha$ we arrive at
\begin{equation*}
F_{ij} = \tfrac1{24} \epsilon_{ijkl} \eIt^\alpha_{mn}
\eIt^\alpha_{mn} F_{kl} + \tfrac12 \sA_{\alpha\beta} \eI^\alpha_{ij}
\eIt^\beta_{mn} F_{mn}~.
\end{equation*}
Using that $\eI^\alpha \cdot \eI^\beta = 4 \delta^{\alpha\beta}$ and
similarly for $\eIt^\alpha$, we see that $\eIt^\alpha \cdot \eIt^\alpha
= 12$, whence
\begin{equation*}
F_{ij} - \tfrac12 \epsilon_{ijkl} F_{kl} = \tfrac12 \sA_{\alpha\beta}
\eI^\alpha_{ij} \eIt^\beta \cdot F~.
\end{equation*}
The left-hand side of this equation is twice the anti-self-dual part of
$F_{ij}$.  Therefore it can be written as a linear combination of the
$\eI^\alpha$.  Contracting with $\eI^\alpha$ we finally arrive at
\begin{equation}\label{eq:Fij}
\eI^\alpha \cdot F = \sA_{\alpha\beta}\, \eIt^\beta \cdot F~.
\end{equation}
It is easy to see that specialising $(ab) = (mn)$ yields the same
equations and that these equations are precisely the first three
equations of \eqref{eq:8instantonfull}.

Now let $(ab) = (im)$ in \eqref{eq:8instantonIJ}.  Only the third term
in the right-hand side contributes now and we find
\begin{equation*}
F_{im} = - \tfrac14 \epsilon_{ijkl} \epsilon_{mnpq} \sA_{\alpha\beta}
\eI^\alpha_{jk} \eIt^\beta_{np} F_{lq}~.
\end{equation*}
Using the anti-self-duality of $\eI^\alpha$ and $\eIt^\alpha$, we
finally arrive at
\begin{equation}\label{eq:Fim}
F_{im} = \sA_{\alpha\beta}\, \eI^\alpha_{ij} F_{jn} \eIt^\beta_{nm}~.
\end{equation}

\section{The local geometry of the moduli space $\eM_K$}
\label{sec:geometry}

In this appendix we review the basic formalism concerning the local
geometry (metric and hyperk\"ahler structure) of the instanton moduli
space $\eM_K$ of a four-dimensional hyperk\"ahler manifold.

On $K$ we have a metric $g_K$ and a chosen hyperk\"ahler structure
consisting of a triplet $\eIt^\alpha$ of complex structures satisfying
the algebra of imaginary quaternions.  The associated K\"ahler
two-forms are anti-self-dual.  Choose a gauge bundle $P$ over $K$ and
let $\eA$ denote the space of connections.  It is foliated by the
group $\eG$ of gauge transformations and the orbit space $\eC \equiv
\eA/\eG$ is the physical configuration space of the gauge theory.  Let
$\eA_+$ denote the space of self-dual connections.  Because
self-duality is a gauge invariant concept, $\eG$ acts on $\eA_+$ and
the orbit space $\eM_K \equiv \eA_+/\eG$ is the moduli space of
instantons.  It is a well-known fact that $\eM_K$ inherits a
hyperk\"ahler metric.  This follows either from an
infinite-dimensional hyperk\"ahler quotient construction as in
\cite{HKLR} or by more pedestrian means (see, e.g., \cite{ItohHK}).
It also follows from supersymmetry as explained in Section
\ref{sec:symhyper}.

Because $K$ is hyperk\"ahler we can reduce the group of the tangent
bundle to $\Sp(1) \subset \SO(4)$.  This means that we can choose
local frames $\mathbf{e}_m$, for $m=1,2,3,4$ which on overlaps are
related via $\Sp(1)$ transformations.  We let $\omega^m$ denote the
dual coframe and will use the shorthand $\omega^{mn\cdots p} =
\omega^m \wedge \omega^n \wedge\cdots\wedge\omega^p$.  Relative to
this frame a connection has components $A_m$.  The metric relative to
this frame is ``flat'' $\langle \mathbf{e}_m,\mathbf{e}_n\rangle =
\delta_{mn}$.  The K\"ahler forms have components $\eIt^\alpha =
\tfrac12 \eIt^\alpha_{mn} \omega^{mn}$.

The study of the local geometry of $\eM \equiv \eM_K$ starts with a
description of its tangent space.  As usual with orbit spaces it is
convenient to think of them as sitting inside the ``top'' space
($\eA_+$ in this case) via a slice.  $\eA_+$ is a subspace of $\eA$
and it is from $\eA$ that it inherits its geometry.  We must therefore
start with $\eA$.

The space $\eA$ of connections is an infinite-dimensional affine
space modelled on the Lie algebra valued one-forms on $K$.  On this
space there is a natural inner product, which induces a metric on
$\eA$:
\begin{equation}\label{eq:metricA}
\langle \delta_1 A, \delta_2 A \rangle = \int_K \Tr \delta_1 A \wedge
\star \delta_2 A = \int_K d^4y\,\sqrt{g_K}\, \Tr \delta_1 A_m \delta_2
A_m~,
\end{equation}
where $\delta_iA$ are Lie algebra valued one-forms.

Now let $A\in\eA_+$ be a self-dual connection.  The tangent vectors
$T_A\eA_+$ to $\eA_+$ at $A$ are precisely those Lie algebra valued
one-forms $\delta A \in T_A\eA$ which obey the linearised self-duality
equations:
\begin{equation}\label{eq:linsd}
D_m\delta A_n - D_n\delta A_m + \epsilon_{mnpq} D_p \delta A_q =
0~.
\end{equation}
In terms of the anti-self-dual K\"ahler forms this is equivalent to
\begin{equation}\label{eq:linsdalt}
\eIt^\alpha_{mn} D_m\delta A_n = 0 \quad\text{for all $\alpha$.}
\end{equation}

We say that a connection is irreducible if there are no nontrivial
solutions $\epsilon$ to $D_m\epsilon=0$.  At an irreducible self-dual
connection $A$, $T_A\eA_+$ breaks up into
\begin{equation}\label{eq:split}
T_A\eA_+ = T_A\eO \oplus T_A\eS~,
\end{equation}
where $\eO \cong \eG$ is the gauge orbit through $A$ and $\eS$ is
the slice at $A$ through the action of the gauge group.  The slice is
locally isomorphic to the moduli space $\eM$ and the map sending a
gauge field $A$ to its gauge equivalence class $[A]$ identifies the
two: $T_A\eS \cong T_{[A]}\eM$.  The tangent space to the gauge orbit
at $A$ can be identified with those Lie algebra valued one-forms $\delta
A$ which take the form of an infinitesimal gauge transformation
$\delta A_m = D_m\epsilon$.  They are automatically tangent to
$\eA_+$---that is, they satisfy the linearised self-duality equation
\eqref{eq:linsd}.  The tangent space to the slice (and by
identification, to the moduli space) are those Lie algebra valued
one-forms $\delta A$ which obey \eqref{eq:linsd} and also which are
orthogonal to the gauge orbits.  This last condition is precisely
Gauss's Law:
\begin{equation}\label{eq:gausslaw}
D_m\delta A_m = 0~.
\end{equation}
The metric on $\eM$ is obtained from the metric on $\eA$ as follows:
given two tangent vectors to $\eM$ at $[A]$ lift them to Lie-algebra
valued one-forms satisfying equations \eqref{eq:linsd} and
\eqref{eq:gausslaw} and then use \eqref{eq:metricA}.  Explicitly, let
$\delta_A A$ for $A=1,2,\ldots,\dim\eM$ be Lie algebra valued one-forms
satisfying \eqref{eq:linsd} and \eqref{eq:gausslaw} and assume that
they define a local frame for $\eM$.  Then the metric relative to this
frame is given by
\begin{equation}\label{eq:instantonmetric}
G_{AB} = \int_K d^4y \sqrt{g_K}\, \Tr \delta_A A_m \delta_B A_m~.
\end{equation}
In other words, the metric on $\eM$ is defined in such a way that the
map taking a self-dual connection to its gauge equivalence class is a
riemannian submersion.

It is possible to show that the metric \eqref{eq:instantonmetric} is
hyperk\"ahler.  In fact, the hyperk\"ahler structure is inherited from
that of $K$.  Define the following two-forms on $\eM$
\begin{equation}\label{eq:kaehlerM}
\eJ^\alpha_{AB} = \int_K d^4y\, \sqrt{g_K}\, \eIt^\alpha_{mn}\,\Tr
\delta_A A_n \delta_B A_m~.
\end{equation} 
It can be shown that $\eJ^\alpha$ define a hyperk\"ahler structure on
$\eM$.

%
%
%
\end{document}